\documentclass[12pt]{iopart}
\usepackage{graphicx}
\usepackage{iopams}

\begin{document}

\title{Controlled formation of metallic nanowires via Au nanoparticle \emph{ac} trapping}

\author{L. Bernard, M. Calame, S. J. van der Molen, J. Liao and C. Sch\"{o}nenberger}

\address{Institute of Physics, University of Basel, 4056 Basel, Switzerland}

\ead{michel.calame@unibas.ch}

\begin{abstract}
Applying \emph{ac} voltages, we trapped gold nanoparticles between micro\-fabricated electrodes
under well-defined conditions. We demonstrate that the nanoparticles can be controllably fused
together to form homogeneous gold nanowires with pre-defined diameters and conductance values.
Whereas electro\-migration is known to form a gap when a \emph{dc} voltage is applied, this
\emph{ac} technique achieves the opposite, thereby completing the toolkit for the fabrication of
nanoscale junctions.
\end{abstract}


\maketitle

\section{Introduction}

Novel nanometer-scale electronic systems have been extensively developed during the last few years.
For example, field-effect transistors made from carbon nano\-tubes~\cite{Wind2002} and organic
molecules~\cite{Liang2002,vanderZant2006} are being designed to constitute alternative routes to
Si-based technology. Molecular electronics in particular combines several advantages:
nanometer-scale size, flexibility of chemical synthesis and discrete energy levels. These aspects
make molecules appear as promising building blocks for ultimate nano-electronics~\cite{heat2003}.
However, a key challenge within this size range remains the control of the connection between a few
molecules and macroscopic current and voltage leads~\cite{hipps2003}. This requires the fabrication
of contacts down to the size of a few nanometers. Electron-beam lithography can presently only
hardly reach \mbox{sub-$10$\,nm} resolution, or under very specific
conditions\cite{Subramanian2004,liu2002,Steinmann2005}. Thus, new techniques need to be developed
to produce nano-scale contacts. Among others, one philosophy is to start from e-beam prefabricated
micro-electrodes and decrease the gap in a second step, for example via
electrodeposition\cite{Morpurgo1999}. A promising method is to close the gap by metallic
nano\-particles, used as inter\-mediates, linking the molecules to the
electrodes~\cite{khondaker2002,Dadosh2005,Long2005,Liao2006,Xu2006}. Hence, there is a need for the
precise positioning of nano\-particles within larger structures.

A simple and controllable method to position nano\-particles is reported here. It consists of
aligning and contacting colloids by means of dielectro\-phoresis (\emph{ac} trapping)
\cite{pohl1978,jones1995,morgan2003}. This technique has previously been used to manipulate a
variety of micro\-meter- and nano\-meter-size objects including bio\-polymers
\cite{washizu1994,dewarrat2002}, cells \cite{pethig1997} and metallic colloids
\cite{Bezryadin1997,Hermanson2001,Amlani2002,khondaker2002,Kretschmer2004,Lumsdon2005}. However, a
systematic approach is still lacking. This is in particular true concerning the assembly of
metallic particles in sub-micrometer gap sizes. In the present work, the process of \emph{ac}
trapping nano\-particles with diameters in between \mbox{$10$\,nm} and \mbox{$100$\,nm} into gaps
of widths ranging from \mbox{$20$\,nm} to \mbox{$500$\,nm} has been investigated. Remarkably, the
technique enables one to form continuous nano\-wires of tunable diameter between microfabricated
electrodes. This process can be initiated by choosing the proper \emph{ac} voltage and series
impedance.

\section{Experimental}

Pairs of planar, tip-shaped gold (Au) electrodes facing each other
were patterned on a Si/SiO$_2$ substrate using conventional UV and
electron-beam lithography (EBL). Before each fabrication step, the
substrate was cleaned in acetone, followed by isopropanol and was
finally exposed to an oxygen plasma for $5$\,min. After resist
spinning, exposure and development, a $5\,$nm thick adhesion layer
of Ti was deposited prior to the evaporation of a $35\,$nm Au
layer. The tips of the electrodes pairs were separated by a gap
typically ranging between $20$ and $100\,$nm. Using an angle
evaporation technique, the gap size can be reduced to less than
$20\,$nm, however with a relatively small yield ($\sim$$5\%$).

Colloidal solutions of charge-stabilised Au particles were
prepared following the method of Turkevich \emph{et al.}
\cite{turkevich1951}. A $0.5\,$mM solution of $HAuCl_4$ was
reduced with aqueous citric acid at boiling temperature. This
resulted in nearly mono\-disperse charge-stabilised Au colloids
with a diameter between $10\,$nm and $30\,$nm, depending on the
reducing agent concentration. Using these charge-stabilised Au
colloids as the starting material, we have also prepared
alkane-thiol encapsulated Au colloids following the method of
Huang \textit{et al.}~\cite{huang2001}. For subsequent
experiments, the encapsulated colloidal particles were dissolved
in a $1:1$ mixture of hexane and dichloromethane.

Trapping of the Au nano\-particles was performed via
dielectro\-phoresis (DEP) by applying an \emph{ac} electric field
$\vec{E}$ between the two electrodes. The dielectro\-phoretic
force $\vec{F}_{DEP}$ acting on a homogeneous, isotropic particle
of radius $a$ is proportional to the gradient of the electric
field amplitude squared and reads \cite{morgan2003}:
\begin{displaymath}
    \vec{F}_{DEP} = \pi a^3 \epsilon_m \cdot
    \Re \left [ \frac{\tilde{\epsilon}_p-\epsilon_m}{\tilde{\epsilon}_p + 2 \epsilon_m} \right ]
    \cdot \vec{\nabla}|\vec{E}|^2
\end{displaymath}
where $\Re$ denotes the real part. The effective polarizability of
the particle in the medium is expressed by the relative
permittivities of the particle $\tilde{\epsilon}_p$ and the medium
$\epsilon_m$. This force typically ranges from \mbox{$0.1$\,pN} to
\mbox{$1$\,pN} \cite{jones1995,morgan2003}.

\begin{figure}[!htb]
  \begin{center}
    \includegraphics[width=8cm]{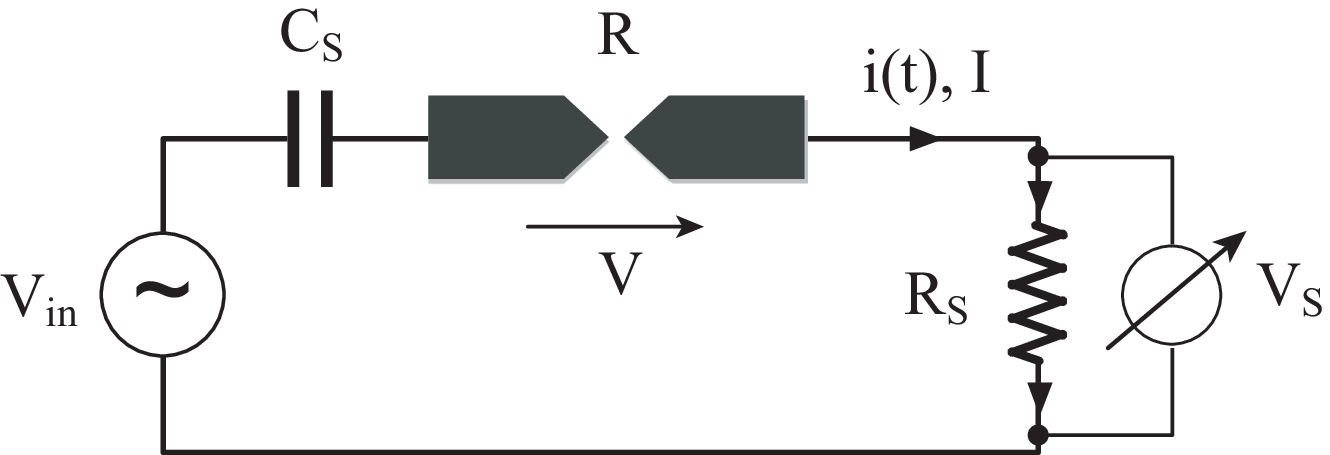}
  \end{center}
\caption{Schematics of the trapping circuit. Both an oscilloscope
and a lock-in amplifier record the voltage drop $V_s$ over the
series resistor $R_s$. Together, $R_s$ and the series capacitor
\mbox{$C_s = 1.5$\,nF} form the total series impedance $Z_s$.}
\label{fig1}
\end{figure}

A circuit diagram of our setup is sketched in Figure~\ref{fig1}. The scheme includes a series
capacitor $C_s$ and a series resistor $R_s$, which constitute a total series impedance $Z_s$. The
capacitor was placed in the circuit to filter out any \emph{dc} component and to avoid
electro\-chemical processes or electro\-migration~\cite{Rodbell1998}. It also prevented further
effects due to, for instance, the electro\-phoretic mobility of the ionic solution, which can
influence the motion of the suspended nano\-particles. To measure the time-dependence of the
current, $i(t)$, we recorded the voltage drop over $R_s$. For this, both a time-averaging lock-in
amplifier and a fast digital oscilloscope were used.  The lock-in amplifier continuously monitored
the trapping process with a time-resolution of \mbox{$\approx 100$\,ms}. In contrast, the
oscilloscope, which provided a much faster time resolution of \mbox{$\approx 100$\,ns}, was
adjusted to trigger at the trapping event, when a significant increase of the current amplitude
$\hat{I}$ was monitored. In this way we measured the delay time until a particle is trapped and the
temporal evolution of the impedance during junction formation. In order to limit the number of
parameters in the process, we fixed the frequency $f$ of the \emph{ac} field throughout the
trapping experiments to a value of 1\,MHz. This allowed the manipulation of nano\-particles with
diameters ranging between $10\,$nm and $100\,$nm at concentrations of order
\mbox{$10^{13}$\,mL$^{-1}$}. Typical values for the total input voltage $V_{in}$ and the series
resistor $R_s$ were \mbox{$V_{in}\leq 3.5$\,V} and \mbox{$0$\,$\Omega \leq R_s \leq
500$\,$\Omega$}, respectively. Throughout this paper, the symbols $I$, $V$, $R$ and $P$ refer to
the RMS values of the current, voltage, resistance and dissipation at (in) the junction,
respectively. During the trapping process, they are time dependent. Their final values, once the
trapping is completed, are denoted by $I_f$, $V_f$, $R_f$ and $P_f$, respectively.

\section{Results and discussion}
\begin{figure}[!htb]
  \begin{center}
    \includegraphics[width=8cm]{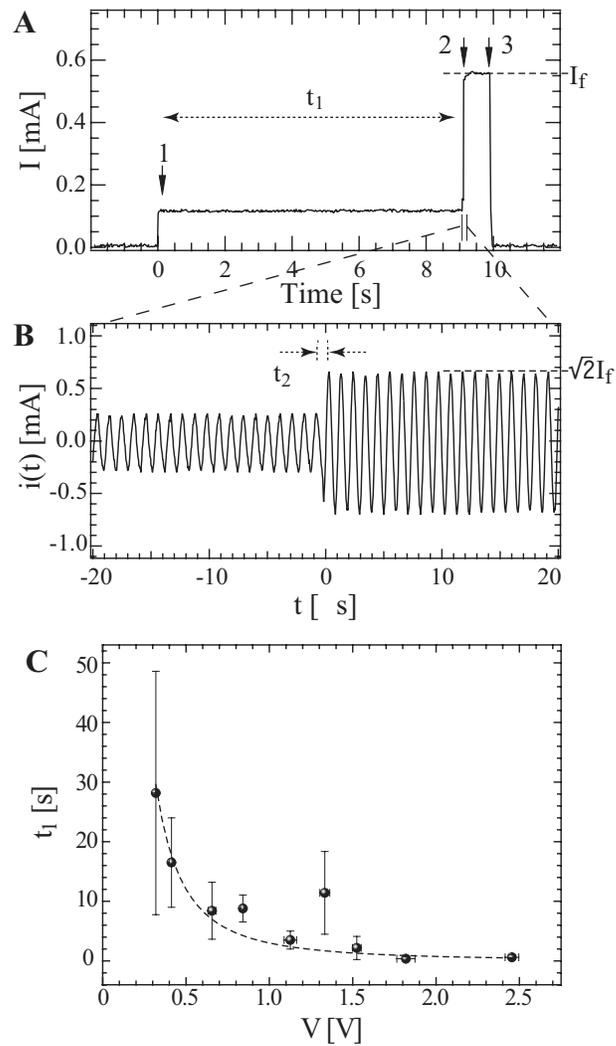}
  \end{center}
\caption{ \textbf{A}. Typical record of the rms current $I$
through the junction during both trapping and anchoring processes,
with \mbox{$V_{in}=1$\,V} and series impedance
\mbox{$|Z_s|=467$\,$\Omega$}. Arrows 1, 2 and 3 mark the time when
the voltage was applied, the contact was made and the voltage was
set to $0$, respectively. The characteristic time for trapping,
$t_1$, is defined as the time between arrows 1 and 2. \textbf{B}.
Close-up view of the \emph{ac} current $i(t)$ measured with an
oscilloscope around the contacting event ($t=0$). \textbf{C}.
$t_1$ as a function of the voltage over the junction $V$ (at
$t<t_1$). Each point is an average over several data points; the
error bars correspond to the standard error. The dotted line
represents a $1/V^2$ dependence.} \label{fig2}
\end{figure}

We explored various types of colloids with a main focus on charge-stabilised Au colloids with a
diameter of $25\,\pm 5$\,nm. To investigate the trapping process in detail, the current through
$103$ junctions was measured during and after the trapping of these specific particles. All samples
were prepared following the same procedure. Figure \ref{fig2}A shows a typical lock-in measurement
of the current $I$ flowing through a junction during trapping. First, a colloidal solution droplet
was deposited on the device ($t<0$). Next, an \textit{ac} trapping voltage $V_{in}$ was applied at
$t=0$ to attract the nano\-particles to the junction. This gave rise to a detectable leakage
current through the solution (arrow 1). After the duration $t_1$ of typically a few seconds to one
minute, a large current increase was suddenly observed (arrow 2). This event signals the formation
of a conducting bridge within the gap due to the trapping of individual nanoparticles. The
conductance jump takes place when the last nanoparticle gets trapped and completes the closing of
the gap. After this increase, the current remained constant and finally, the input voltage $V_{in}$
was set to zero (arrow 3). We note that in a few particular cases the current did not increase in a
single step, but rather displayed multiple steps, suggesting the formation of a few conducting
bridges in parallel. Figure \ref{fig2}B shows a typical oscilloscope trace of $i(t)$, set up to
trigger exactly at the current increase (arrow 2, Figure \ref{fig2}A). We see that the junction
changes its resistance typically within \mbox{$t_2\sim 1\,\mu$s}. Figure \ref{fig2}C shows the
dependence of the average trapping time $t_1$ on the voltage over the junction $V$ (for $t<t_1$).
The trapping time roughly scales with the inverse of the voltage squared $1/V^2$, which agrees with
the expectation for a driven particle movement impeded by viscous friction. In that case the
velocity scales with the driving force which is, according to the expression for the
dielectrophoretic force, proportional to the electric field squared, and hence to the voltage
squared.

\begin{figure}[!htb]
  \begin{center}
    \includegraphics[width=16cm]{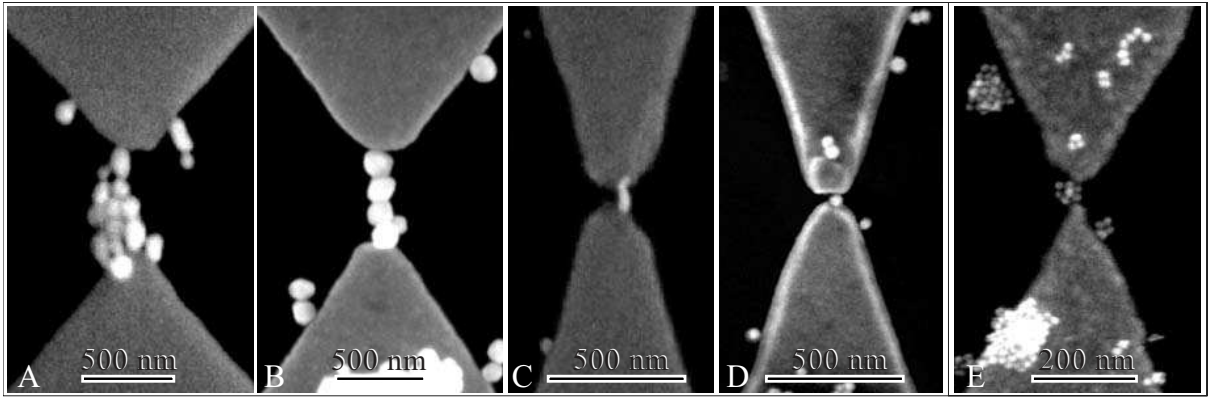}
  \end{center}
\caption{Scanning electron micrographs of junctions prepared under
various conditions. \textbf{A}. Large gap ($500\,$nm) with large
nano\-particles ($120 \pm 20\,$nm)~: parameters $V_{in}=2$\,V,
$|Z_s|=111\,\Omega$; final resistance $R_f=280\,\Omega$.
\textbf{B}. Large gap (560\,nm) with large nano\-particles ($120
\pm 20\,$nm)~: $V_{in}=1$\,V, $|Z_s|=111\,\Omega$;
$R_f=50\,\Omega$. \textbf{C}. Small gap (40\,nm) with a wire
formed by small nano\-particles ($25 \pm 5\,$nm)~: $V_{in}=1$\,V,
$|Z_s|=148\,\Omega$; $R_f=160\,\Omega$. \textbf{D}. Small gap
(30\,nm) with a single small nano\-particle ($25 \pm 5$\,nm)~:
$V_{in}=1.8$\,V, $|Z_s|=467\,\Omega$; $R_f=4.8$\,M$\Omega$.
\textbf{E}. Small gap (20\,nm) with small $C_{12}$-functionalized
nano\-particles ($10 \pm 1$\,nm)~: $V_{in}=1$\,V,
$|Z_s|=147\,\Omega$; $R_f=4$\,M$\Omega$.} \label{fig3}
\end{figure}

After each trapping experiment, scanning-electron microscopy (SEM)
inspection of the samples revealed that in $67$ cases the
junctions showed no sign of breakdown. Figure \ref{fig3} reports a
set of representative results. First, we focus on Figures
\ref{fig3}A and \ref{fig3}B. Here, nano\-particles with a
relatively large diameter of $120 \pm 20$\,nm were trapped into
$\sim$ 500\,nm large gaps. The trapping conditions for both cases
were identical, except for the applied voltage $V_{in}$:
\mbox{$2$\,V} in case A and \mbox{$1$\,V} in case B. We clearly
observe a larger number of trapped particles in A than in B. This
illustrates that one can tune the number of nano\-particles
bridging the gap by adjusting the trapping voltage. The electric
field required for optimal trapping amounts to about $E\approx
10^7$\,V/m. In Figure \ref{fig3}B, a well-ordered chain of
nano\-particles was formed with a junction resistance of
\emph{only} $50\,\Omega$. This is surprising and indicates that
after the nano\-particles were trapped, a second, `anchoring'
process took place. During anchoring, the colloids become
physically and electrically connected, leading to a low final
resistance. Figure \ref{fig3}C shows a device in which even
smaller particles, i.e. $25\,\pm 5$\,nm, were trapped, with an
applied voltage \mbox{$V_{in}=1$\,V}. Remarkably, the individual
colloids are not distinguishable any longer. Instead of building a
chain, as in Figure \ref{fig3}B, the particles have fused together
forming a \emph{wire} with a well defined diameter. The final
resistance of this wire is comparatively low, i.e. $160\,\Omega$.
Figure \ref{fig3} shows that the anchoring process can lead to the
formation of two distinct structures: chains or wires. Below, we
will discuss this phenomenon in more detail. Finally, in Figure
\ref{fig3}D and E, we show examples of junctions with relatively
large final resistances in the M$\Omega$ range. In Figure
\ref{fig3}D, a single nanoparticle ($25\,\pm 5$\,nm) was trapped.
In contrast to Figures \ref{fig3}A-D where charge-stabilised
colloids were used, Figure \ref{fig3}E shows a typical result for
the trapping of dodecanethiol ($C_{12}H_{25}S$)-functionalized
nano\-particles of \mbox{$10$\,nm} diameter within a
\mbox{$30$\,nm} gap. Unlike the former cases, devices made with
functionalized particles never displayed fusing and the resistance
values were always large ($> M\Omega$).

\begin{figure}[!htb]
  \begin{center}
    \includegraphics[width=8cm]{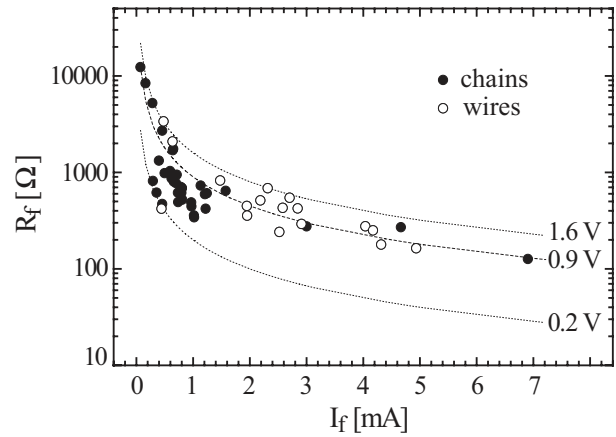}
  \end{center}
\caption{ Final resistance $R_f$ versus current $I_f$ of the
junctions showing successful trapping and anchoring (56 devices).
The voltage over the junctions $V_f=I_f R_f$ is bound by
\mbox{$0.2$\,V} from below (no trapping) and \mbox{$1.6$\,V} from
above (breakdown after trapping), as indicated by dotted lines. A
typical voltage is \mbox{$0.9$\,V} (dashed line). Solid dots:
chain formation; open dots: wire formation (as extracted from SEM
images).} \label{fig4}
\end{figure}

In $11$ cases of the $67$ successful trapping experiments with \mbox{$25\,\pm 5$\,nm} diameter
charge-stabilised nano\-particles, the final junction resistance $R_f$ was much larger than
$1$\,M$\Omega$. This indicates that no metallic contact was established between the nano\-particles
and the electrodes. An overview of the rest of the data set ($56$ experiments) is given in Figure
\ref{fig4}. Here, we plot the final device resistance $R_f$ as a function of the current $I_f$,
both measured at the end of the trapping and anchoring process. There is quite some scatter in the
final resistance of the devices. Nevertheless, the data set is confined between a lower and an
upper bound given by the voltage $V_f=I_f R_f$. More precisely, the voltage over the junction
always lies in between \mbox{$0.2$\,V} and \mbox{$1.6$\,V}, as indicated by the dotted lines, a
typical value being \mbox{$0.9$\,V}. The lower bound in Figure \ref{fig4} is related to the
trapping process: if the electric field is too small, trapping does not take place. In contrast, a
voltage that is too large leads to sample destruction after trapping as evidenced by `burnt'
electrodes which are modified on a macroscopic scale. This breakdown process is likely related to a
thermal run-away by excessive Joule heating as reported in electro\-migration experiments on
nano\-junctions~\cite{Durkan1999,Lambert2003,Trouwborst2006}. The open and solid data points in
Figure~\ref{fig4} relate to two different kinds of junctions. The solid dots refer to junctions in
which the nano\-particles formed chains during the trapping process (cf. Figure~\ref{fig3}B),
whereas the open dots relate to wire formation (cf. Figure~\ref{fig3}C). It appears that the two
subsets group together in Figure~\ref{fig4}, despite the clear scatter. Hence, Figure~\ref{fig4}
forms a good basis to discuss about the possible processes behind anchoring, chain and wire
formation, as well as device breakdown.

\begin{figure}[!htb]
  \begin{center}
    \includegraphics[width=8cm]{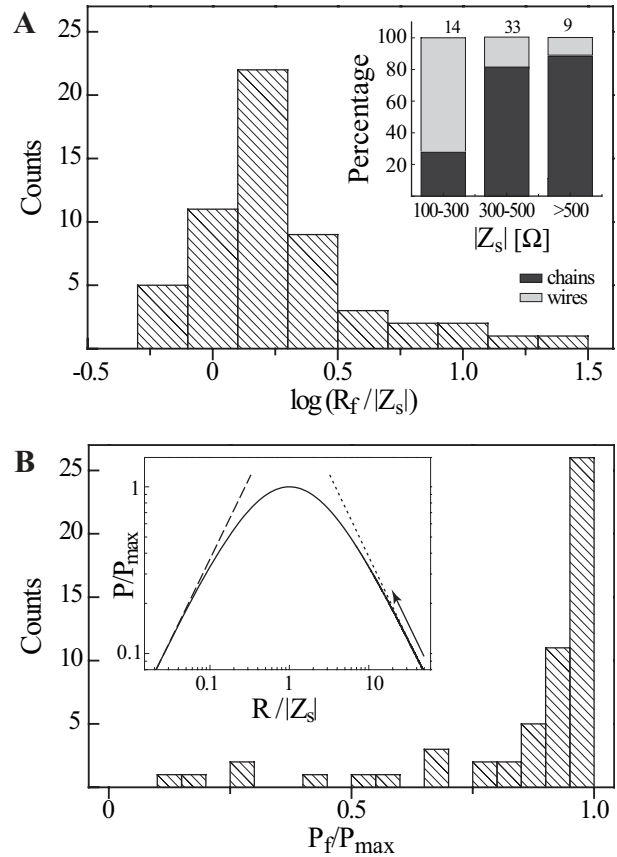}
  \end{center}
\caption{ \textbf{A}. Histogram of $log(R_f/|Z_s|)$, for the data
in Figure~\ref{fig4}. A single peak is seen with a maximum around
$log(R_f/|Z_s|)=0.2$. The inset shows the percentage of chains
(dark) and wires (light) for different values of $|Z_s|$. Wires
are more favorably formed for lower $|Z_s|$. \textbf{B}. Histogram
of $P_f/P_{max}$, where $P_f=V_f^2/R_f$. A clear majority of the
devices has $P_f \approx P_{max}$. The inset shows a calculation
of $P/P_{max}$ as a function of $R/|Z_s|$, which reaches its
maximum at $R=|Z_s|$ (Calculation for $Z_s=452-134i \Omega$). The
arrow indicates that the resistance of our junctions decreases
during homogenisation. The dashed and dotted lines refer to the
limiting cases of current- and voltage biasing, respectively.}
\label{fig5}
\end{figure}

We first focus on the moment when a set of nano\-particles is
being trapped in the junction at $t=t_1$. These particles
experience a large electric field in the gap region \mbox{$\sim
10^7$\,V/m}, which induces a dipole in each of them. Consequently,
the dipole-dipole interaction between the particles becomes
significant, tending to line up the particles in the junction.
This explains the preference for the initial chain formation. At
that point, the charge-stabilised particles are not yet
inter\-connected, so that the whole voltage drops over the tiny
gaps ($\leq$1\,nm) between the particles. This has two
consequences. First, the electric field is much more localised
than before chain formation. Hence, the range of the DEP force is
decreased, leading to a much smaller trapping probability for
left-over particles in the solution. In this way, DEP is
self-limiting. Second, the field over the gaps between the
individual colloids tend to become very high, \mbox{$\geq
10^8$\,V/m}. At such fields, surface diffusion of Au atoms is
enhanced in the direction of the gap and can lead to the formation
of narrow Au bridges between the
nano\-particles~\cite{Tsong1991,Mendez1996}. If no connections are
formed, a device will have a high final resistance ($\gg$ 1
\,M$\Omega$), as observed in $11$ of our $71$ junctions made with
charge-stabilised particles. As mentioned above, junctions
prepared with alkanethiol-covered particles \textit{always}
yielded high resistance values,  suggesting that the alkanethiol
shell protects the colloids against field-driven atomic migration.
However, in most devices with charge-stabilised particles, the
particles were inter\-connected and the resistance of the devices
dropped dramatically. One may wonder what determines the final
value $R_f$ of these junctions. To address this point, we refer to
Figure \ref{fig5}A, where a histogram of $log(R_f/|Z_s|)$ is
plotted. This graph describes the probability distribution of the
ratio of $R_f$ to the total series impedance $|Z_s|$. We find that
$R_f$ is typically of the same order of magnitude as $|Z_s|$. The
histogram features a clear peak with a maximum around a ratio of
$log(R_f/|Z_s|)=0.2$, corresponding to $R_f=1.6|Z_s|$. In the
inset of Figure \ref{fig5}A, we show a stacked bar plot of the
percentage of chains (dark) and wires (light) for three ranges of
$|Z_s|$. Interestingly, chains are more common for higher $|Z_s|$,
while wires are mostly found for lower $|Z_s|$. Figure \ref{fig5}A
indicates that we can tune (at least roughly) the device
properties via the series impedance.

To understand this, we bring forward a possible model of wire
formation. We first note that a change in shape from chain to wire
leads to a decrease of the total surface energy. Hence, we expect
that surface tension is the effective driving force for the fusing
process. Also, in order to transform a chain of spherical
particles into a rod-like wire, a considerable diffusion of Au
atoms is needed. Since diffusion is an activated process, an
increase of the junction temperature by Joule heating will clearly
facilitate wire formation. Before fusing, $R \gg |Z_s|$ and the
junction is voltage-biased with the power $P$ dissipated over the
junction given by $P=V_{in}^2/R$. Once atomic diffusion sets in
and the gaps start to fill up, the resistance decreases and
consequently the dissipation increases. If the applied voltage
$V_{in}$ is too large, this can lead to a thermal run-away and
sample breakdown. However, the series impedance $Z_s$ also has an
effect on $P$. It can self-limit the thermal run-away process for
moderate $V_{in}$. To see this, we note that the junction would be
current-driven in the opposite regime $R << |Z_s|$. In that case,
the local dissipation $P$ is given by $P\approx RI^2$ and
decreases with decreasing $R$, thus limiting the fusion process.
The full dependence of $P$ on $R/|Z_s|$ is shown in the inset of
Figure~\ref{fig5}B. There is a maximum power
$P_{max}=V_{in}^2/4|Z_s|$ at $R=|Z_s|$. The dashed and dotted
lines indicate the dissipation in the limiting cases of
effectively current-biased and voltage-biased junctions,
respectively. Based on this consideration, we expect the process
to stop when $R \approx Z_s$ where the dissipation $P$ reaches its
maximum. Hence, the final dissipation $P_f$, measured at the end
of successful junction formation, should be close to $P_{max}$. In
Figure~\ref{fig5}B, we display a histogram of $P_f/P_{max}$ for
all data points in Figure~\ref{fig4}. For the vast majority of our
devices, Figure~\ref{fig5}B shows that $P_f\approx P_{max}$, as
anticipated. Because the final resistance $R_f$ is generally a bit
larger than $|Z_s|$ \mbox{(Figure~\ref{fig5}A)}, we infer that the
process tends to stops to the right of the maximum in the inset of
Figure~\ref{fig5}B.

Relating the junction temperature $T$ to the power $P$ is not straightforward, but it can be stated
that $T$ is related to the volume power density $p$. If we assume for simplicity that only the
junction cross-section $A$ changes during wire formation, but not its length $L$, we have
$p:=P/AL$. Because $R\propto 1/A$ one may also write $p\propto PR$. Hence, in the beginning, when
$R \gg |Z_s|$ and $P=V_{in}^2/R$, the power density is constant, even if $A$ increases due to the
thermally driven diffusion of Au atoms. This homo\-ge\-nization process can therefore go on at
constant temperature until the assumption $R \gg |Z_s|$ is no longer valid. Then, $p$ and hence $T$
will decrease, limiting atomic diffusion. This picture suggests that the homogenisation process can
proceed further for small $|Z_s|$ as compared to large $|Z_s|$. It therefore becomes understandable
why wire-like devices appear for small $|Z_s|$, whereas chain-like ones appear for large $|Z_s|$,
consistent with the inset of Figure~\ref{fig5}A.

In the previous discussion, electromigration was ignored, since it does not occur for the \emph{ac}
current densities we apply \cite{Rodbell1998}. In fact, electromigration at similar \emph{dc}
current densities leads to gap formation~\cite{Liang2002,vanderZant2006,Trouwborst2006}, which is
the \emph{opposite} effect compared to the new `gap-closing' process that we report here.
Interestingly, the series impedance and Joule heating play a similar role in electro\-migration
experiments as in the work presented here~\cite{Trouwborst2006}. We speculate that repeated gap
formation and closing should be possible by using \emph{ac} and \emph{dc} voltages, alternately.

\section{Conclusions}
We demonstrate a simple method to control the formation of metallic nano\-wires and chains of
desired length and diameter. It is based on the dielectro\-phoretic trapping of nano\-particles. We
point out the relevance of two processes: the trapping and the subsequent anchoring of the
particles between the electrodes. By choosing a proper series impedance, one can tune the final
resistance and appearance of a junction: lower series impedances favor wire formation, whereas
higher series impedances result in chains. The nano\-structures produced are promising elements in
the framework of nano\-electronics. For instance, they can be used to contact nano\-meter-size
building blocks, possibly in combination with \emph{dc} electro\-migration. Using this approach, it
may become possible to fine tune nano\-gaps in electro\-migration devices, which is of high
interest to molecular electronics.

\begin{ack} We are grateful to M. Steinacher for his technical support. S.J.vd M. acknowledges the
Netherlands Organisation for Scientific Research, NWO ('Talent stipendium'). This work was
supported by the Swiss National Science Foundation, the Swiss National Center of Competence in
Research "Nanoscale Science" and the European Science Foundation through the Eurocore program on
Self-Organised Nanostructures (SONS).
\end{ack}

\section*{References}
\bibliographystyle{unsrt}
\bibliography{coltrap}

\newpage

\clearpage

\end{document}